\documentclass{moriond}

\usepackage{amsmath,amsfonts}
\usepackage{xspace}
\usepackage{multirow}
\usepackage[square,sort,comma,numbers]{natbib}
\usepackage{float}
\usepackage{tabularx}
\usepackage[justification=raggedright]{caption}
\usepackage[flushleft]{threeparttable}
\usepackage{slashed}
\usepackage{indentfirst}
\usepackage{xcolor}
\usepackage{ulem}

\def\Journal#1#2#3#4{{#1} {\bf #2}, #3 (#4)}

\begin{document}
\vspace*{4cm}
\title{Review of Nucleon Decay Searches at Super-Kamiokande}

\author{Volodymyr Takhistov \\
(for the Super-Kamiokande Collaboration)}

\address{Department of Physics and Astronomy, University of California, Irvine, CA 92697, USA}

\maketitle\abstracts{Baryon number violation appears in many contexts.
It is a requirement for baryogenesis and is a consequence of Grand Unified Theories (GUTs),
which predict nucleon decay. Nucleon decay searches provide the most
direct way to test baryon number conservation and also serve as a unique
probe of GUT scale physics around $10^{14-16}$ GeV.
Such energies cannot be reached directly by accelerators. 
However, they can be explored indirectly at large 
underground water Cherenkov (WC) 
experiments, which due to the size of
their fiducial volume are highly sensitive to nucleon decays.
We review searches for baryon number violating
processes at the state of the art WC detector, the Super-Kamiokande.  Analyses of
the typically dominant non-SUSY and SUSY nucleon decay channels such as 
$p \rightarrow (e^+, \mu^+) \pi^0$ and $p \rightarrow \nu K^+$,
 as well as more exotic searches, will be discussed. 
Presented studies set the world's best limits, 
which circumvent the allowed parameter space of theoretical models.}

\section{Motivation}

Baryon number $(B)$ is a global accidental symmetry of the Standard Model (SM)
and ensures that the lightest baryon (proton) is stable. However,
not only is it violated by non-perturbative effects within the SM itself \cite{'tHooft:1976up},
baryon number violation ($\slashed{B}$) is a required condition
for explaining the observed matter - anti-matter asymmetry of the universe \cite{Sakharov:1967dj}.
More so, various theoretical arguments as well as reductionism hint at the SM being an incomplete theory
and that there exists a more unifying underlying account. Such an account could be provided
by Grand Unified Theories (GUTs) 
\cite{Pati:1973uk,Georgi:1974sy, Fritzsch:1974nn}, like $SU(5)$ and $SO(10)$,
 which unite the three SM gauge groups together
$G_{GUT} \supset SU(3)_C \otimes SU(2)_W \otimes U(1)_Y$
and offer an explanation for charge quantization as well as coupling unification.
With quarks and leptons appearing within a common GUT representation,
one can transform into another, giving rise to nucleon decay (explicit $\slashed{B}$).
In the more fundamental domain of quantum gravity,
it is expected \cite{Banks:2010zn} that global symmetries, such as $B$, are violated in general.
The presence of nucleon decay could also have
profound consequences for the future fate of the universe \cite{Adams:1996xe}.
For a topical review of $B$-violation as well as nucleon decay predictions
see Ref.~\cite{Babu:2013jba} and
Ref.~\cite{Nath:2006ut}, respectively.

While there is no convincing evidence of $B$-violation thus far, testing $B$ remains a high priority.
Nucleon decay can provide not only one of the most striking signatures of $B$-violation,
but it also offers a unique way of testing Grand Unified Theories.
With the unification scale being around $10^{15 \pm 1}$ GeV,
Grand Unified Theories cannot be probed directly by accelerators.  
However, they can be examined indirectly with 
large underground water Cherenkov experiments, which
due to the size of their fiducial volume are highly sensitive to nucleon decays.
Below, we review the nucleon decay search results from  
the current state of the art WC experiment, the Super-Kamiokande (SK, Super-K).

\section{The Super-Kamiokande Experiment}

Super-Kamiokande is a 50 kiloton WC detector (22.5 kiloton fiducial
volume) located beneath a one-km rock overburden (2700 m.~water equivalent)
within the Kamioka mine in Japan. The SK detector
is composed of an inner (11,146 inward-facing 20-inch PMTs, providing
40\% photo-coverage) and an outer (1,855 8-inch outward-facing PMTs) detector, which are optically separated.
Cherenkov radiation \cite{Frank:1937fk}, produced by charged particles traveling through
water, is collected by the PMTs and is used to reconstruct the physics events. 

Data collected by Super-Kamiokande (SK) during the periods of
SK-I (May~1996-Jul.~2001, 1489.2 live days), 
SK-II \footnote{An accident caused SK-II photo-coverage
 to be reduced to $20\%$. Full photo-coverage was restored in SK-III.} (Jan.~2003-
Oct.~2005, 798.6 live days), SK-III (Sep.~2006-Aug.~2008,
518.1 live days) and the ongoing SK-IV 
\footnote{Electronics have been upgraded in SK-IV.} experiment 
(Sep.~2008-present, $>1500$ live days) corresponds to a combined
exposure of more than 250 kiloton$\cdot$yrs.
Details of the detector design, performance, calibration, data reduction and
simulation can be found in Ref.~\cite{Fukuda:2002uc, Abe:2013gga}.

The experiment allows one to study a 
wide variety of physics topics in the MeV - TeV energy range,
including those related to solar and
 atmospheric neutrinos (oscillations, tests of Lorentz invariance, day-night asymmetry,
sterile neutrino, indirect dark matter searches), 
supernovae relic neutrinos as well as nucleon decay.

\section{Nucleon Decay Searches}

For nucleon decay analyses, only events for which all of the observed
Cherenkov light was fully contained within the inner detector are considered.
The observed Cherenkov rings are classified 
either as showering (``$e$-like", for $e^{\pm}$ and $\gamma$) 
or non-showering (``$\mu$-like'', for $\mu^{\pm}$). We do not distinguish
between signal channels with a $\nu$ and $\overline{\nu}$ in the final state, since neither 
of the neutrinos is observable.

In the signal Monte Carlo (MC) simulations, the effects of Fermi momentum,
nuclear binding energy as well as nucleon-nucleon
correlated decays are taken into account \cite{Nishino:2012bnw}.
The atmospheric neutrino background interactions are
generated using the neutrino flux calculations
 of Honda \textit{et. al.} \cite{Honda:2006qj} and the
NEUT simulation package~\cite{Hayato:2002sd}, which uses a relativistic
Fermi gas model. The SK detector simulation software is
based on the GEANT-3 \cite{Brun:1994aa} package. Background MC
corresponding to a 500 year detector exposure-equivalent is
generated for each SK period.

\subsection{$p \rightarrow e^+ \pi^0$ and $p \rightarrow \mu^+ \pi^0$}

The $p \rightarrow e^+\pi^0$ channel is often the most dominant nucleon decay mode in GUTs,
with typical lifetime predictions of $10^{29 - 36}$ yrs.
Previous searches for this channel
have already excluded minimal $SU(5)$ 
\cite{Haines:1986yf, Gajewski:1989gh, Hirata:1989kn, Shiozawa:1998si}.
Within some models (e.g. flipped $SU(5)$ \cite{Ellis:2002vk}), 
a similar channel, $p \rightarrow \mu^+ \pi^0$,
 can also appear with a significant branching ratio.
\begin{figure}[htb]
\begin{minipage}{0.47\linewidth}
\centerline{\includegraphics[width=1\linewidth]{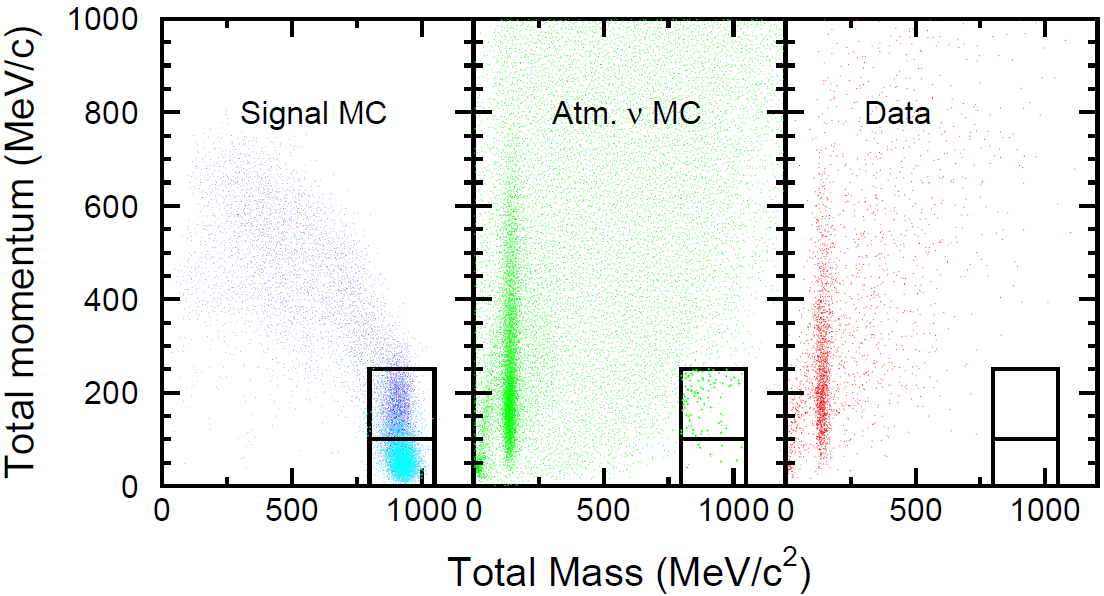}}
\end{minipage}
\hfill
\begin{minipage}{0.47\linewidth}
\centerline{\includegraphics[width=1\linewidth]{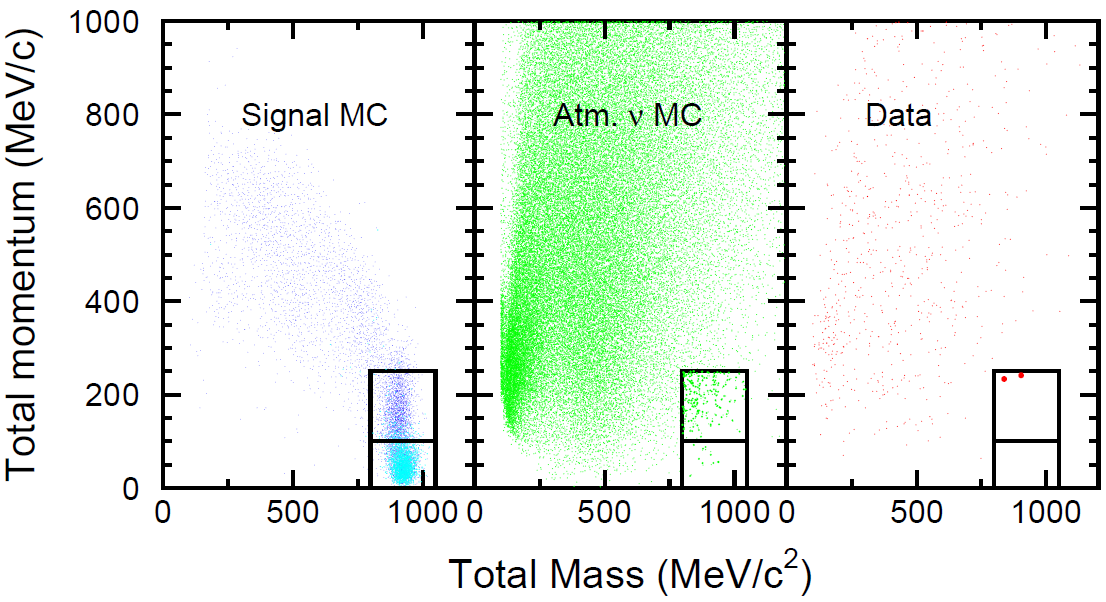}}
\end{minipage}
\caption[]{Reconstructed invariant mass vs. total momentum
for $p \rightarrow e^+\pi^0$ [left] and $p \rightarrow \mu^+\pi^0$ [right].
The left panels show signal MC, where light blue corresponds to free protons and dark
blue to bound protons. The middle panels show atmospheric-$\nu$ MC
and the right panels show data. From Ref.~\cite{skprep}.}
\label{fig:results_p2emupi0}
\end{figure}

Since $e^+$, $\mu^+ (\rightarrow e^+\nu\nu)$ 
as well as $\pi^0 (\rightarrow \gamma\gamma)$
produce visible Cherenkov rings, one can fully
 reconstruct the invariant mass and momentum 
of the parent proton. Figure~\ref{fig:results_p2emupi0} 
displays the signal MC, background MC and 
data (306 kiloton$\cdot$yrs of exposure),
after all the event selection criteria have been 
applied. The signal region consists of two portions,
a ``lower box" (free protons) and an ``upper box" 
(bound protons), separated in the analysis for improved sensitivity.
For $p \rightarrow e^+ \pi^0$, the average 
signal efficiency as well as the total expected background
within the selected region is 38.7\% and 0.61 events, respectively. 
For $p \rightarrow \mu^+ \pi^0$, it is 34.6\% and 0.87 events, respectively.
No data events pass the selection for $p \rightarrow e^+ \pi^0$, 
while two events pass for $p \rightarrow \mu^+\pi^0$.
The Poisson probability of observing 
two such events for a given exposure is $23\%$.  Since
both events also display background-like 
features, they are judged as coming from atmospheric-$\nu$ background. 
Hence, the 90\% confidence level (C.L.) lower 
lifetime limits of $1.7 \times 10^{34}$ yrs. and $7.8 \times 10^{33}$ yrs. are placed
on the $p \rightarrow e^+\pi^0$ and 
$p \rightarrow \mu^+\pi^0$ channels \cite{skprep}, respectively.

The analyses for the other $p \rightarrow l^+ + m^0$ 
and $n \rightarrow l^- + m^+$  searches are conducted in a similar manner.
Here, $l^{\pm} (e^{\pm}, \mu^{\pm})$ represents a 
charged lepton, while $m^0 (\omega^0, \eta, \rho^0, K^0)$ 
and $m^+ (\pi^-, \rho^-)$ denote a neutral
 and a charged meson, respectively. No significant signal excess
is observed in any of these channels, resulting in the 90 \% C.L.  
lower lifetime limits of around $10^{32 - 33}$ years \cite{Nishino:2012bnw, Regis:2012sn}.

\subsection{$p \rightarrow \nu K^+$}

The introduction of supersymmetry (SUSY), which is theoretically well motivated, 
allows one to make GUT coupling unification precise.
New operators come into play and typically the dominant SUSY GUT mode is
$p \rightarrow \nu K^+$, with lifetime predictions 
around $10^{29 - 36}$ yrs. Previous searches for this channel
have already excluded minimal (TeV-)SUSY $SU(5)$ \cite{Kobayashi:2005pe}.
\begin{figure}[htb]
\begin{minipage}{0.33\linewidth}
\centerline{\includegraphics[width=0.95\linewidth]{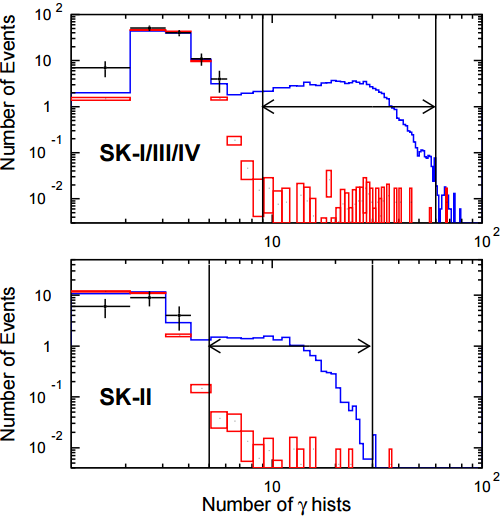}}
\end{minipage}
\hfill
\begin{minipage}{0.32\linewidth}
\vspace*{0.1cm}
\centerline{\includegraphics[width=0.98\linewidth]{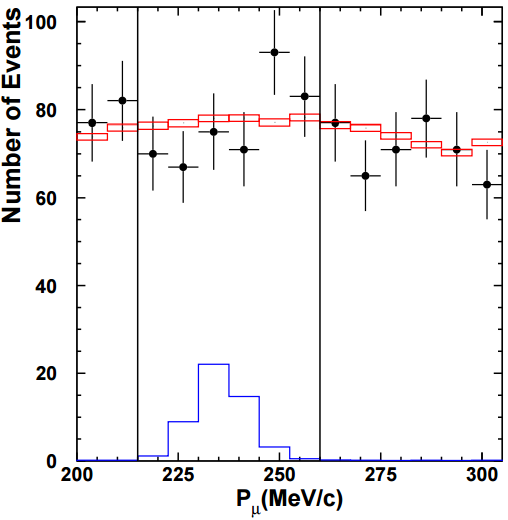}}
\end{minipage}
\hfill
\begin{minipage}{0.32\linewidth}
\vspace*{0.1cm}
\centerline{\includegraphics[width=0.97\linewidth]{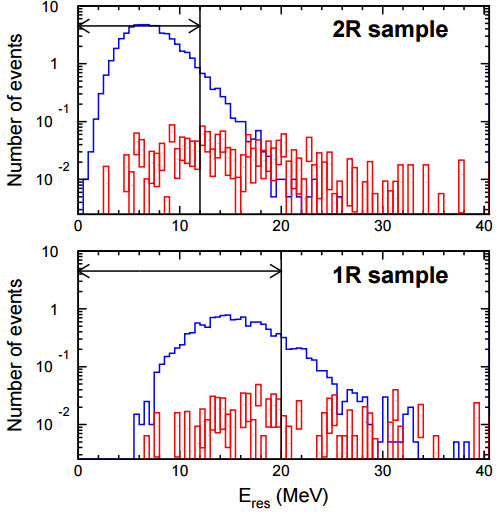}}
\end{minipage}
\caption[]{Results for three different $p \rightarrow \nu K^+$ search methods.
promt-$\gamma$ tagging [left], $\mu^+$ momentum [center] and 
visible energy for $\pi^+$ [right]. The signal MC, atmospheric $\nu$ MC and
data are shown in blue, red, and black, respectively. From Ref.~\cite{Abe:2014mwa}.}
\label{fig:results_p2nuK}
\end{figure}
Since neutrino and the kaon are invisible 
(kaon below Cherenkov threshold) in this mode,
reconstruction of the parent nucleon is not possible. However, 
analysis can be done on the kaon decay products,
coming from the $K \rightarrow \mu^+\nu~(Br.~64\%)$ 
and $K \rightarrow \pi^+\pi^0~(Br.~21\%)$ channels.
Three different analyses are performed at SK. 
The first search is done in the muon channel, 
where a prompt-$\gamma$, which could appear
 from nuclear de-excitation after the proton has decayed, is tagged.
The second search is also done in the muon channel, where 
a spectral $\chi^2$ fit to the $\mu^+$ momentum is performed.
The third search is done in the pion channel, 
where the $\pi^0 (\rightarrow \gamma\gamma)$ is reconstructed and the
visible energy from $\pi^+$ is analyzed. 
Figure~\ref{fig:results_p2nuK} displays the signal MC, 
background MC and data (260 kiloton$\cdot$yrs exposure),
after all the event selection criteria have been applied. 
The average efficiency and background 
for the three searches is 7.9\% and 0.39 events, 
33.7\% and 579.4 events and 8.2\% and
 0.56 events for the prompt-$\gamma$, $\mu$-spectrum and
the pion search, respectively. No significant excess is observed and 
a 90\% C.L. lower lifetime limit of $6.6 \times 10^{33}$ yrs. 
is placed on this channel \cite{Abe:2014mwa}.

\subsection{$n - \overline{n}$}

For this mode, $\Delta B = 2$ and it parametrizes
 the scale of $U(1)_{B-L}$ symmetry breaking.
The process is naturally connected with baryogenesis as well as neutrinos, 
since Majorana neutrinos and the see-saw mechanism require $\Delta L = 2$.
Since the effective operator
corresponding to this process has dimension nine, 
$n - \overline{n}$ can be viewed as a probe of 
intermediate $10^{3} - 10^{11}$ GeV scale physics \cite{Mohapatra:2009wp}.
\begin{figure}[htb]
\begin{minipage}{0.33\linewidth}
\centerline{\includegraphics[width=1\linewidth]{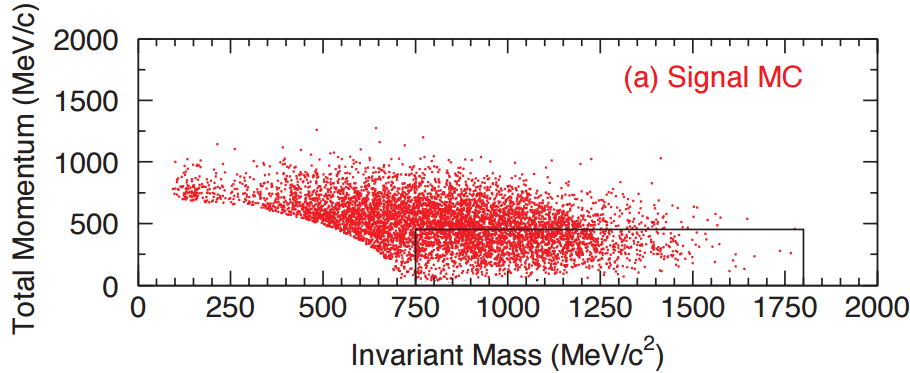}}
\end{minipage}
\hfill
\begin{minipage}{0.32\linewidth}
\vspace*{0.1cm}
\centerline{\includegraphics[width=1\linewidth]{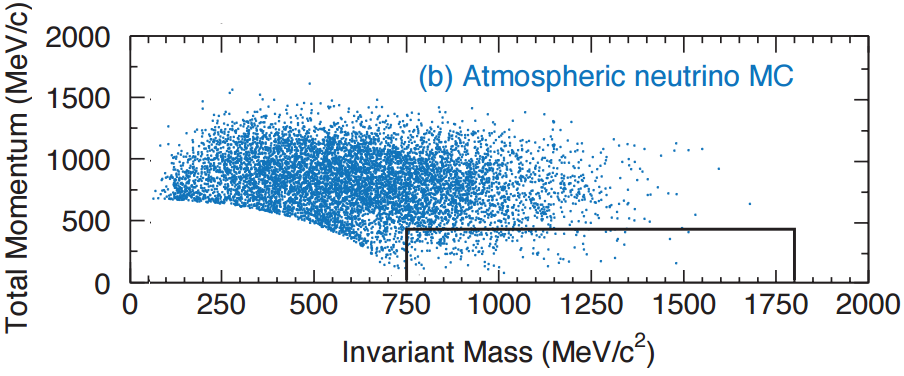}}
\end{minipage}
\hfill
\begin{minipage}{0.32\linewidth}
\vspace*{0.1cm}
\centerline{\includegraphics[width=1\linewidth]{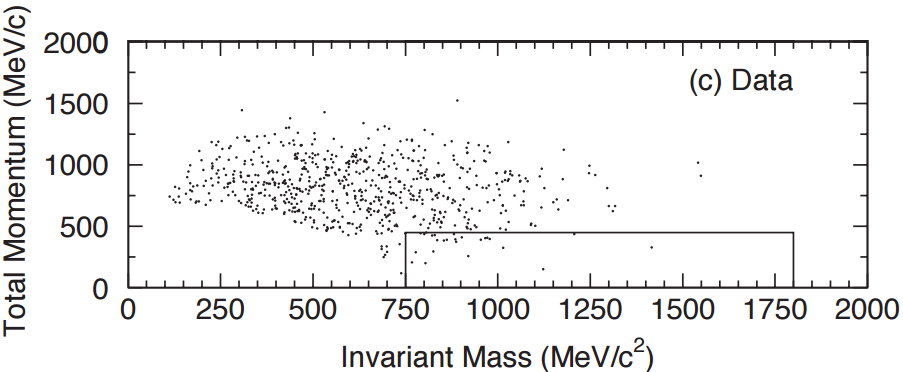}}
\end{minipage}
\caption[]{Reconstructed invariant mass vs. total momentum
for $n - \overline{n}$.
The signal MC, atmospheric $\nu$ MC and
data are displayed on the left (red), center (blue) and right (black) panels, respectively.
From Ref.~\cite{Abe:2011ky}.}
\label{fig:results_nnbar}
\end{figure}
When the resulting $\overline{n}$ is captured by a $p$ or $n$,
a variety of mesons (predominantly pions) are released. 
This can be used to reconstruct the 
invariant mass and momentum of the original
``di-nucleon" $\overline{n}p$ or $\overline{n}n$ systems and perform 
a search analysis similar to $p \rightarrow e^+ \pi^0$.
Figure~\ref{fig:results_nnbar} displays the signal MC, 
background MC and data (91.5 kiloton$\cdot$yrs exposure),
after all the event selection criteria have been applied. 
The efficiency and expected background for this search is 12.1\% and 24 events, respectively.
No significant excess is observed and 
a 90\% C.L. lower lifetime limit of $1.9 \times 10^{32}$ 
yrs.\footnote{This result can  be converted into a 
limit for $n - \overline{n}$ oscillations in vacuum, yielding lifetime of around 
$10^8$ s.} is obtained \cite{Abe:2011ky}.

\subsection{$np, nn, pp \rightarrow mesons$}

Similar to $n - \overline{n}$, di-nucleon decays of $np, nn$ and $pp$
are $\Delta B = 2$ processes and can provide
novel insights beyond the single nucleon decay searches.
The di-nucleon decay channels
$np \rightarrow \pi^+\pi^0, nn \rightarrow \pi^0\pi^0$
and $pp \rightarrow \pi^+\pi^+$
allow testing models with an extended Higgs sector and
suppressed proton decay \cite{Arnold:2013cva}.
Within the context of supersymmetry, the
di-nucleon decay mode $pp \rightarrow K^+K^+$ 
can provide the most sensitive experimental probe
of the $R$-parity violating coupling 
$\lambda_{112}''$ \cite{Barbieri:1985ty,Goity:1994dq}.
Due to complicated systematics, associated with 
the final state mesons and their nuclear interactions, 
as well as the resulting convoluted multi-ring signatures,
the analyses for the di-nucleon decays to mesons are done
using a boosted decision tree (BDT) \cite{Hocker:2007ht}.
This allows one to get a far better discrimination between signal and background
than by solely applying the event selection criteria. 
We illustrate this with a $pp \rightarrow K^+K^+$ search in SK-I \cite{Litos:2014fxa}.

\begin{figure}[htb]
\centering
\includegraphics[width=0.45\linewidth]{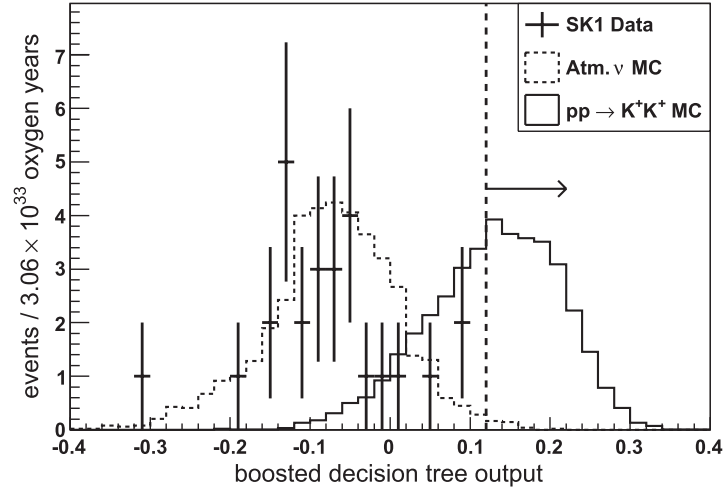}
\caption[]{Final output of the boosted decision tree for $pp \rightarrow K^+K^+$ search.
Events to the right of the cut at 0.12 are considered 
signal candidates. From Ref.~\cite{Litos:2014fxa}.}
\label{fig:dinuceKK}
\end{figure}

Unlike $p \rightarrow \nu K^+$, the kaons in $pp \rightarrow K^+ K^+$ 
are above Cherenkov threshold and produce visible rings.
Subsequently, the kaons decay, primarily through the
$K^+ \rightarrow \mu^+\nu, \pi^+\pi^0$ channels.
The two resulting kaon vertices are spatially separated by $\sim2$ meters.
After the pre-selection criteria have been applied, the remaining events
are processed with BDT. The final BDT output is displayed in Figure~\ref{fig:dinuceKK}.
The cut on the final BDT output variable is chosen
at 0.12, maximizing
\noindent the $(\text{Sig.}/\sqrt{\text{Sig.} + \text{Bkg.}})$ ratio.
Events to the right of the cut are taken to be signal candidates.
With (without) BDT, one achieves the final signal efficiency of 12.6\% (21.9\%) and an expected
total SK-I background of 0.3 (33.9) events. 
The data is found to be consistent with background
and a 90\% C.L. lower lifetime limit of $1.7 \times 10^{32}$ yrs.\footnote{The lifetime 
limits for di-nucleon decays are calculated per $^{16}$O nucleus.} is placed.
This lifetime limit can be translated \cite{Goity:1994dq} into a limit 
on the $\lambda_{112}''$ SUSY $R$-parity violating coupling,
resulting in $|\lambda_{112}''| < 7.8 \times 10^{-9}$ \cite{Litos:2014fxa}.

Similarly, no significant signal excesses are observed in the di-nucleon
$np \rightarrow \pi^+\pi^0, nn \rightarrow \pi^0\pi^0$
and $pp \rightarrow \pi^+\pi^+$ decay searches, 
resulting in the 90\% C.L. lower lifetime limits
of around $10^{32}$ yrs. for all three modes \cite{Gustafson:2015qyo}.

\subsection{Spectral Searches}

Due to a large theoretical uncertainty in the nucleon
decay predictions, it is important to study
a variety of channels. Since $e$-like and 
$\mu$-like single-ring atmospheric-$\nu$ background is well understood,
an array of modes producing such signatures can be readily analyzed.
These include $p \rightarrow (e^+,\mu^+)\nu\nu, 
p \rightarrow (e^+,\mu^+)X$ \footnote{Here, $X$ denotes a massless and invisible particle.},
$np \rightarrow (e^+,\mu^+, \tau^+) \nu$ and $n \rightarrow \nu\gamma$ channels.
The trilepton $p \rightarrow (e^+,\mu^+)\nu\nu$ decays
 can appear within Pati-Salam models, with some \cite{Pati:1983jk} predicting
their lifetime to be around $10^{30-33}$ yrs. 

\begin{figure}[htb]
 \hspace{-5em}
\vspace{0.6em}
\begin{minipage}[b]{0.4\textwidth}
\centering
 \hspace{-2.5em}
\includegraphics[width=0.95\textwidth]{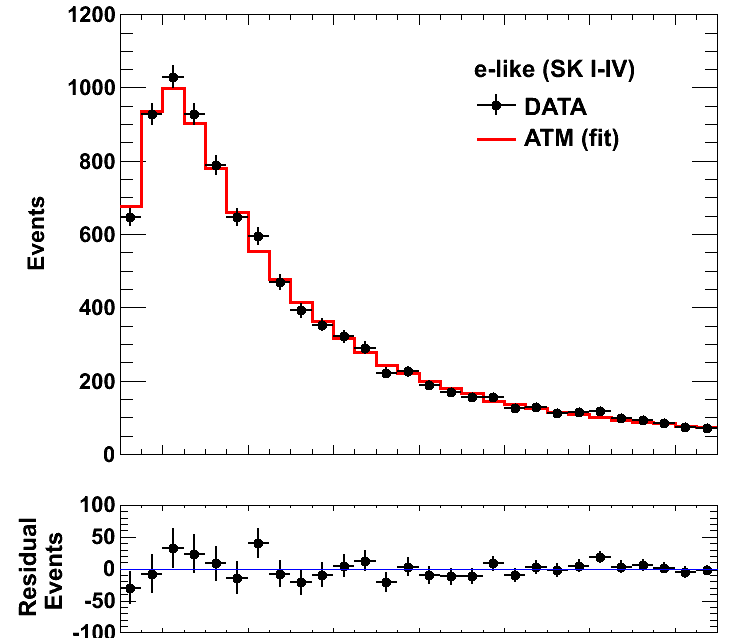}
\vspace{0.2em}
\end{minipage}
\begin{minipage}[b]{0.4\textwidth}
\centering
\hspace{-2.75em}
\vspace{0.1em}
\includegraphics[width=0.945\textwidth]{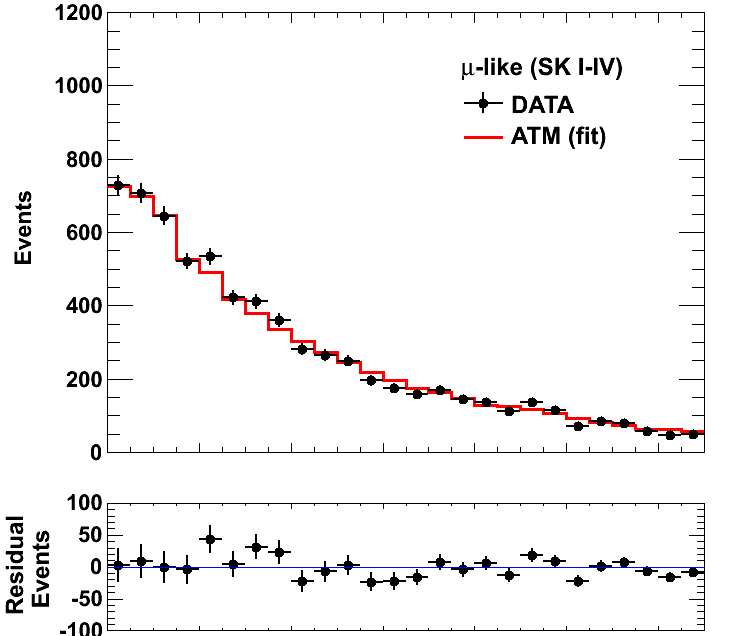}
\end{minipage}
\begin{minipage}[b]{0.4\textwidth}
\centering
\includegraphics[width=0.995\textwidth]{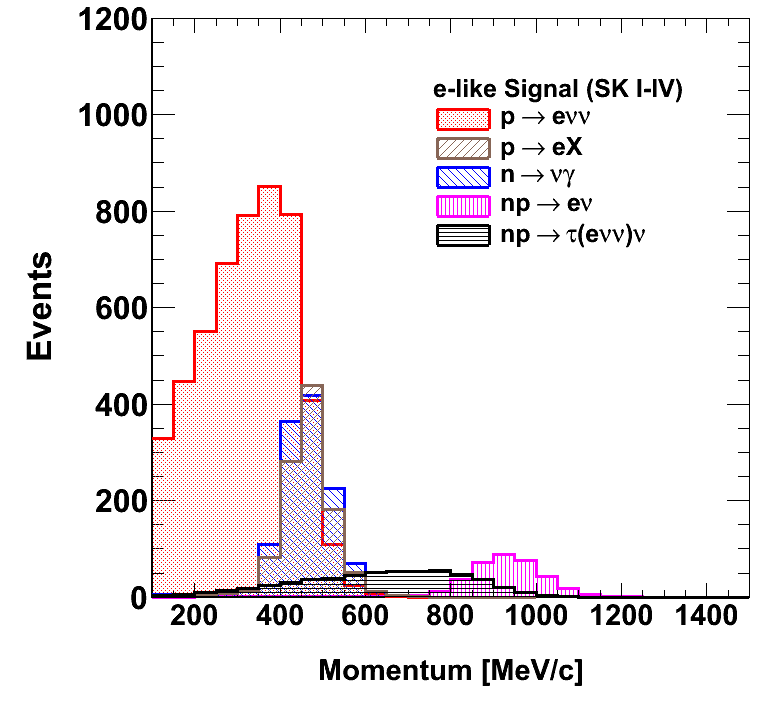}
\end{minipage}
\hspace{6em}
\begin{minipage}[b]{0.4\textwidth}
\centering
\includegraphics[width=0.98\textwidth]{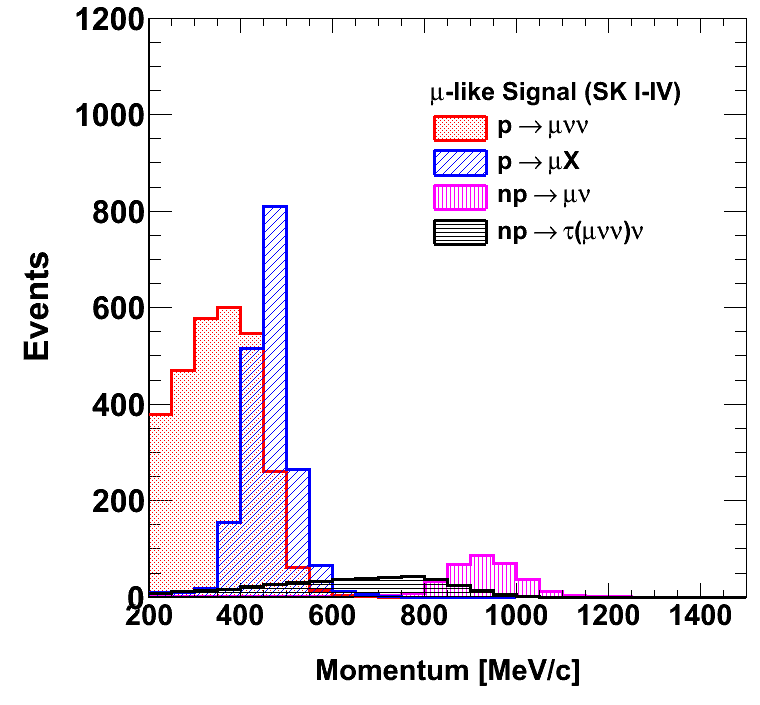}
\vspace{-0.05em}
\end{minipage}
\caption{[top] Reconstructed momentum distribution, with the SK data 
(black dots) and the best-fit result for the 
atmospheric-$\nu$ background MC (solid line) displayed.
The corresponding residuals are shown below, after the 
fitted background has been subtracted from the data.
[bottom] The 90\% confidence level allowed
 nucleon decay signal multiplied by 10 (hatched histograms),
from the signal and background MC fit to data. All modes are shown (overlaid),
with $e$-like channels on the left and $\mu$-like channels on the right. 
From Ref.~\cite{Takhistov:2014pfw,Takhistov:2015fao}.}
\label{fig:spectral_full}
\end{figure}


The search for all these channels is performed 
by applying a spectral ``pull-method" $\chi^2$ fit \cite{Fogli:2002pt}
to the $e$-like and the $\mu$-like single-ring momenta, after the event selection.
Figure~\ref{fig:spectral_full} displays the best-fit result for the 
atmospheric-$\nu$ background MC, data (273.4 kiloton$\cdot$yrs exposure) and
the 90\% C.L. allowed amount of nucleon decay signal
($\times 10$) after the fit.
For these searches, the average signal efficiency is $\sim 95\%$ for the $e$-like modes 
and $\sim 80\%$ for the $\mu$-like modes, with the latter being lower due to the 
decay-electron detection efficiency \footnote{The decay-electron detection
efficiency is improved in SK-IV by 20\% compared to other SK periods, due to upgraded electronics.}.
With no significant excess observed in either search, 
90\% C.L. lower lifetime limits of around $10^{32}$ 
yrs. are placed on all of the above channels \cite{Takhistov:2014pfw,Takhistov:2015fao}.

\begin{table}[htb]
\caption[]{Summary of Super-Kamiokande nucleon decay search results.}
\label{tab:fullresults}
\vspace{0.4cm}
\begin{center}
\begin{threeparttable}[htb]
\begin{tabular}{|l|c|c|c|}
\hline
& & & \\
Decay Mode  & $|\Delta (B - L)|$ & $\tau/{\mathcal{B}}~(90\% C.L.)$ &  Reference \\
\hline
$p \rightarrow e^+\pi^0$  			& $0$ 		& $1.7 \times 10^{34}$~yrs. &  \cite{skprep} \\ 
$p \rightarrow \mu^+\pi^0$  			& $0$ 		& $7.8 \times 10^{33}$~yrs. &  \cite{skprep} \\ 
$p \rightarrow \nu K^+$	  			& $0(\overline{\nu}), 2(\nu)$ & $6.6 \times 10^{33}$~yrs. & \cite{Abe:2014mwa}  \\ 
$p \rightarrow \mu^+ K^0$	  		& $0(\overline{\nu}), 2(\nu)$ & $6.6 \times 10^{33}$~yrs. & \cite{Regis:2012sn}  \\ 
$p \rightarrow e^+\eta$	  			& $0$		 & $4.2 \times 10^{33}$~yrs. &  \cite{Nishino:2012bnw} \\ 
$p \rightarrow \mu^+\eta$			& $0$		 & $1.3 \times 10^{33}$~yrs. &  \cite{Nishino:2012bnw} \\ 
$p \rightarrow e^+\rho^0$			& $0$		 & $7.1 \times 10^{32}$~yrs. &  \cite{Nishino:2012bnw} \\ 
$p \rightarrow \mu^+\rho^0$			& $0$		 & $1.6 \times 10^{32}$~yrs. &  \cite{Nishino:2012bnw} \\ 
$p \rightarrow e^+\omega^0$		& $0$		 & $3.2 \times 10^{32}$~yrs. &  \cite{Nishino:2012bnw} \\ 
$p \rightarrow \mu^+\omega^0$		& $0$		 & $7.8 \times 10^{32}$~yrs. &  \cite{Nishino:2012bnw} \\
$p \rightarrow \nu\pi^+$				& $0(\overline{\nu}), 2(\nu)$		 & $3.9 \times 10^{32}$~yrs. &  \cite{Abe:2013lua}  \\
$p \rightarrow e^+\nu\nu$			& $0(\overline{\nu}\nu), 2 (\nu\nu, \overline{\nu}\overline{\nu})$		
									& $1.7 \times 10^{32}$~yrs. &  \cite{Takhistov:2014pfw}  \\
$p \rightarrow \mu^+\nu\nu$	& $0(\overline{\nu}\nu), 2 (\nu\nu, \overline{\nu}\overline{\nu})$		
									& $2.2 \times 10^{32}$~yrs. &  \cite{Takhistov:2014pfw}  \\
$p \rightarrow e^+X$\tnote{a}		& $0 (X?)$	 & $7.9 \times 10^{32}$~yrs. &  \cite{Takhistov:2015fao} \\
$p \rightarrow \mu^+X$\tnote{a}		& $0 (X?)$	 & $4.1 \times 10^{32}$~yrs. &  \cite{Takhistov:2015fao} \\
$n \rightarrow e^+\pi^-$				& $0$		 & $2.0 \times 10^{33}$~yrs. &  \cite{Nishino:2012bnw} \\ 
$n \rightarrow \mu^+\pi^-$			& $0$		 & $1.0 \times 10^{33}$~yrs. &  \cite{Nishino:2012bnw} \\ 
$n \rightarrow e^+\rho^-$			& $0$		 & $7.0 \times 10^{31}$~yrs. &  \cite{Nishino:2012bnw} \\ 
$n \rightarrow \mu^+\rho^-$			& $0$		 & $3.6 \times 10^{31}$~yrs. &  \cite{Nishino:2012bnw} \\
$n \rightarrow \nu\pi^0$				& $0(\overline{\nu}), 2(\nu)$		 & $1.1 \times 10^{33}$~yrs. &  \cite{Abe:2013lua}  \\
$n \rightarrow \nu\gamma$			& $0(\overline{\nu}), 2(\nu)$		 & $5.5 \times 10^{32}$~yrs. &  \cite{Takhistov:2015fao} \\
$pp \rightarrow K^+K^+$			& $2$		 & $1.7 \times 10^{32}$~yrs.\tnote{b} &  \cite{Litos:2014fxa}  \\
$pp \rightarrow \pi^+\pi^+$			& $2$		 & $7.2 \times 10^{31}$~yrs.\tnote{b} &  \cite{Gustafson:2015qyo} \\
$np \rightarrow e^+\nu$				& $0(\overline{\nu}), 2(\nu)$		 & $2.6 \times 10^{32}$~yrs.\tnote{b} &  \cite{Takhistov:2015fao} \\
$np \rightarrow \mu^+\nu$			& $0(\overline{\nu}), 2(\nu)$		 & $2.0 \times 10^{32}$~yrs.\tnote{b} &  \cite{Takhistov:2015fao} \\
$np \rightarrow \tau^+\nu$			& $0(\overline{\nu}), 2(\nu)$		 & $3.0 \times 10^{31}$~yrs.\tnote{b} &  \cite{Takhistov:2015fao} \\
$np \rightarrow \pi^+\pi^0$			& $2$		 & $1.7 \times 10^{32}$~yrs.\tnote{b} &  \cite{Gustafson:2015qyo} \\
$nn \rightarrow \pi^0\pi^0$			& $2$		 & $4.0 \times 10^{32}$~yrs.\tnote{b} &  \cite{Gustafson:2015qyo} \\
$n - \overline{n}$ oscillations			& $2$		 & $1.9 \times 10^{32}$~yrs. &  \cite{Abe:2011ky} \\
\hline
\end{tabular}
\begin{tablenotes}[flushleft]\footnotesize
 \item [a] $X$ denotes a single particle which is assumed to be massless and invisible. 
 \item [b] The lifetime limits for di-nucleon decays are calculated per $^{16}$O nucleus. 
\end{tablenotes}
\end{threeparttable}
\end{center}
\end{table}

\section{Summary and Outlook}

Baryon number violation is motivated by various theoretical considerations
and can be tested by nucleon decay searches. Large underground water Cherenkov
detectors are highly sensitive to nucleon decays. We have reported 
on results for an extensive array of nucleon decay searches carried out at the 
current state of the art WC experiment, Super-Kamiokande.
No significant signal excess has been found in any of the analyzed channels.
These results, which correspond to the world's
 best lifetime limits, are summarized in Table~\ref{tab:fullresults}.
The majority of the limits are already in the $10^{32}$ yrs. range, signifying that even
some of the more baroque types of theoretical models are constrained. The projected
gadolinium dissolution in the SK detector will allow us to further 
reduce the nucleon decay background and thus improve on the
search sensitivity. Future experiments,
such as Hyper-Kamiokande and DUNE/LBNF, are expected 
to improve the presented results by up to an order of magnitude.
Since the majority of the 
nucleon lifetime predictions from the various theoretical models
lie below the $\sim 10^{36}$ yrs. range, significant
discovery potential is expected in the upcoming searches.


\section*{Acknowledgments}

The author would like to thank the Moriond EW-2016 conference organizers for
the opportunity to present these results. The Super-Kamiokande
collaboration gratefully acknowledges the cooperation of the
Kamioka Mining and Smelting Company. The
experiment was built and has been operated
with funding from the Japanese Ministry of Education,
Culture, Sports, Science and Technology, the U.S.
Department of Energy, and the U.S. National Science
Foundation.


\end{document}